\documentclass{article}
\usepackage{graphicx}        % For eps figures, newer & more powerfull
\usepackage{color}           % For color text: \color command
\usepackage{hyperref}
\usepackage{breakurl}        % For breaking URLs easily trough lines
\usepackage{natbib}
\newcommand{\etal}{{\it et al.}}

\begin{document}
\sloppy

%%%%%%%%%%%%%%%%%%%%%%%%%%%%%%%%%%%%%%%%%%%%%%%%%%%
\title{Cyclic changes of the photospheric magnetic \\field polarity}

\author{E.~S.~Vernova; M.~I.~Tyasto\\
IZMIRAN SPb Filial, St.-Petersburg, Russia\\
elenavernova96@gmail.com\\
D.~G.~Baranov\\
Ioffe Institute, St.-Peterburg, Russia
}
\date{}
%%%\author{\firstname{M.~I.}~\surname{Tyasto}}
\maketitle
%%%%%%%%%%%%%%%%%%%%%%%%%%%%%%%%%%%%%%%%%%%%%%%%%%%

\begin{abstract}
The distribution of magnetic fields of positive and negative polarities over the surface of the Sun was studied. Synoptic maps of the photospheric magnetic field (NSO Kitt Peak, 1978--2016) were used for the analysis. In the time-latitude diagram for weak magnetic fields ($B \leq 5$\,G), inclined bands are clearly visible, indicating the alternating dominance of magnetic fields with positive or negative polarity drifting towards the poles of the Sun.
Analysis of the time-latitude diagram using the method of empirical orthogonal functions (EOF) made it possible to establish a cyclic change in the polarity of magnetic field flows with a period of 1-3 years, which indicates a possible connection with quasi biennial variation. Similar period values are obtained by averaging over the latitude of the time-latitude diagram
\end{abstract}
%\end{opening}
%-------------------------------------------------

\section{Introduction}
     \label{S-Introduction}
The magnetic fields responsible for the emergence and development of all types of solar activity are distributed over the Sun in a complex way depending on the strength of the field and the phase of the solar cycle. The transport of magnetic fields in the photosphere (surges) plays an important role in the development of the solar activity cycle. The influence of different magnetic field flows is determined by their scales and life time. A large number of works are devoted to the study of these phenomena. One important result of these studies is the discovery of a close association of polar field reversal with the transport of magnetic fields to the pole by the so-called ``Rush-to-the-Poles'' (RTTP) flows.
This phenomenon has been studied in the green coronal line \citep{alt}, in solar filaments  \citep{gopa}, in photospheric magnetic fields \citep{petr} and  \citep{mord}. These flows, drifting from latitudes $\sim 40^\circ$ to poles and causing a change in the sign of the polar field, are the product of the decay of the following sunspots. Each of these flows is dominated by the magnetic field of one pronounced polarity, namely the polarity of the following sunspots, which is opposite to the sign of the polar field.
But there are also known flows of fields with an alternating sign of the dominant field, which follow one after another with a period of 1--2 years. These flows found by \citep{vecc}, were considered as one of manifestations of the  quasi biennial variations (QBO).The properties of such flows (or “ripples”)  were described in detail in \citep{ulri}. The reasons for the appearance of such flows are still unclear. It was noted \citep{petr} that the problem of isolating their source is very difficult, since there may be several sources. Attempts have been made to establish a link between flows and decaying active regions \citep{mord}. Other authors \citep{ulri} believed that the cause of such flows was not related to active regions. Apparently, the question of the origin of alternating polarity flows remains unresolved and requires further research.
Weak magnetic fields and their distribution on the surface of the Sun are of great interest, since they occupy a significant proportion of the solar surface and play an important role in the development of solar cycle models. As shown by our calculations based on NSO Kitt Peak data, fields with $|B| \leq 5$\,G cover $65\%$ of the solar surface and fields with $5$\,G$< |B| \leq 10$\,G – $18\% $; in total $83\%$ of the  Sun’s surface are occupied by fields $|B|\leq 10$\,G.The spatiotemporal evolution of weak photospheric magnetic fields was studied in (\citet{murs} and references therein). In our work, we examine the flows with the different polarities from data on weak magnetic fields over 4 solar cycles.
%%%%%%%%%%%%%%%%%%%%%%%%%%%%%%%%%%%%%%%%%%%%%%%%%%%%%%%%%%%%%%%%%%

\section{Data and Method}
\label{KP-Data}
Based on synoptic maps of the photospheric magnetic field of the NSO Kitt Peak Observatory (1978--2016), the distribution of magnetic fields of positive and negative polarity over the surface of the Sun was studied. Synoptic maps had a resolution of $1^\circ$ in longitude (360 steps) and 180 steps in sine latitude from $-1$ (southern pole) to $+1$ (northern pole). Each map consisted of $360\times180$ pixels containing the magnetic field values in Gauss. Synoptic maps averaged over longitude taking into account the sign of the magnetic field were used in the construction of the time-latitude diagram. When analyzing the spatiotemporal properties of the solar magnetic fields, especially weak fields ($|B|< 5$\,G), the possible effects of random fluctuations in field strength (noise) should be taken into account. In the polar regions of the Sun, the noise \citep{harv} reaches 2\,G per pixel on NSO Kitt Peak synoptic maps (the greatest noise is observed at the poles). When building a time-latitude diagram for each of the synoptic maps 360 longitude values are averaged. Due to this averaging  the influence of random factors is largely reduced. In our case, we can consider as an error the value of 0.1\,G, that is, the smallest significant number in the table representation of synoptic maps.
As a rule, in the time-latitude diagram, strong fields in the form of Maunder butterflies are the most prominent ones, and the distribution of weak fields is poorly distinguished. Since we were interested in the role of weak fields in the distribution of the magnetic field on the surface of the Sun, we limit the influence of strong fields. For this purpose, a saturation threshold of 5\,G was set for each synoptic map. As a result, only fields modulo less than 5\,G ($|B|\leq 5$\,G) were left unchanged on each synoptic map, while larger or smaller fields were replaced by the corresponding limit values  of $+5$\,G or  $-5$\,G. Maps converted in this way were used to build a time-latitude diagram.

%%%%%%%%%%%%%%%%%%%%%%%%%%%%%%%%%%%%%%%%%%%%%%%%%%%%%%%%%%Figure1
\begin{figure}[t]
   \centerline{\includegraphics[width=0.95\textwidth,clip=]{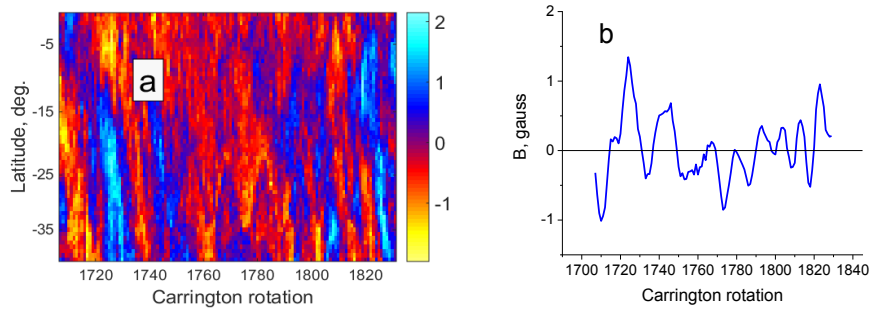}
              }
              \caption{Cyclic variations of the magnetic field polarity of the photosphere. (a)~A section of the time-latitude diagram for the southern hemisphere (latitudes from $0^\circ$ to $-40^\circ$, time period -- Carrington rotations 1700--1840). Bands of blue and red correspond to flows dominated by a positive or negative field.
(b)~Same time period. Time variation of the magnetic field at latitude $-33^\circ$ of the southern hemisphere. The variation is smoothed by 5 points with adjacent averaging along the time axis. The maxima and minima of the curve correspond to positive and negative magnetic flows.
                      }
   \label{mapprof}
   \end{figure}
%%%%%%%%%%%%%%%%%%%%%%%%%%%%%%%%%%%%%%%%%%

%%%%%%%%%%%%%%%%%%%%%%%%%%%%%%%%%%%%%%%%%%%%%%%%%%%%%%%%%%

\section{Results and Discussion}
\label{Results}
As an  example, Figure~\ref{mapprof}a shows a section of a time-latitude diagram (southern hemisphere, latitudes  from $0^\circ$ to $-40^\circ$, time period -- Carrington rotations 1700--1840). Blue and red bands correspond to the flows dominated by a positive or negative field. The width of the bands is about one year. The inclination of these bands on the diagram suggests that in time the flows drift in latitude in the direction from the equator to the poles. These flows are different from Rush-to-the-Poles (RTTP). RTTP flows can be seen as  wide bands with a width of about 2--3 years, which begin at latitudes $\sim 30^\circ -40^\circ$ and have constantly the same polarity as the  following sunspots. The arrival of RTTP to the poles leads to the polar field reversal.
Unlike RTTP, the flows we are considering are narrower (width 0.5--1 yr); they appear near the equator and have an alternating dominant polarity. Following the term used in \citep{ulri} we will call such cyclic variations of magnetic field ``ripples''. We observe ripples in the period of time during which the sign of the polar field is constant (between two RTTPs). This result differs from the conclusions of the work of \citet{vecc}, the authors of which see similar structures only in years of high solar activity. On the other hand \citet{ulri} concluded that ripples are present constantly, regardless of the level of solar activity.
%%%%%%%%%%%%%%%%%%%%%%%%%%%%%%%%%%%%%%%%%%%%%%%%%%%%%%%%%%Figure2
\begin{figure}[t] 
   \centerline{\includegraphics[width=0.95\textwidth,clip=]{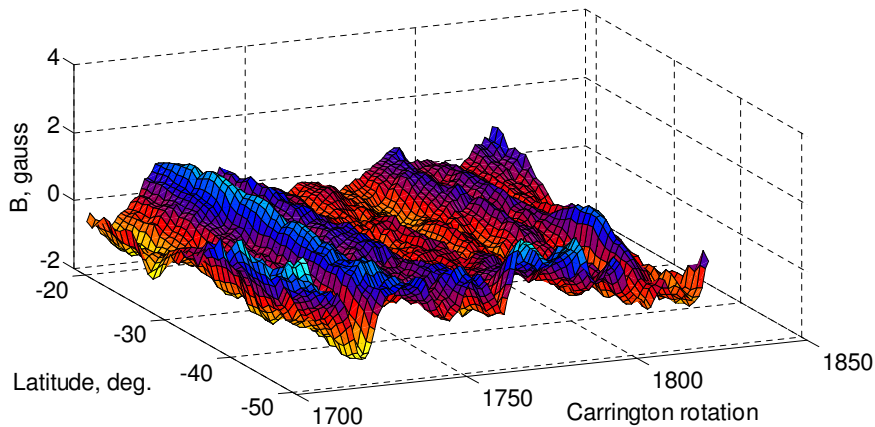}
              }
              \caption{ 
3D representation of the time-latitude diagram (same diagram as in Figure~\ref{mapprof}a). The diagram is smoothed by 5 points with adjacent averaging along the time axis. The alternation of positive and negative polarities of the magnetic field is clearly visible as ridges and valleys in the map. 
                      }
   \label{pamir}
   \end{figure}
	%%%%%%%%%%%%%%%%%%%%%%%%%%%%%%%%%%%%%%%%%%%%%%%%%%%%%%
Figure~\ref{mapprof}b shows how the magnetic field varies over time at a fixed latitude ($-33^\circ$ of the southern hemisphere). The maxima and minima of the curve in Figure~\ref{mapprof}b correspond to positive and negative magnetic flows. Alternation of the magnetic field flows of opposite polarities (ripples) can be seen. Variations in polarity are observed for 9 years from 1981 to 1990, i.e. include a phase of decline in solar activity, a minimum, and a phase of rise.
%%%%%%%%%%%%%%%%%%%%%%%%%%%%%%%%%%%%%%%%%%%%%%%%%%%%%%%%%%%%%%%%%%%%%%%%%%%%%%%%%%%%%Figure3
\begin{figure}[t] 
   \centerline{\includegraphics[width=0.95\textwidth,clip=]{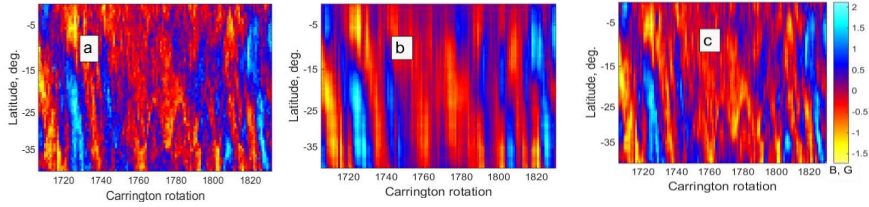}
              }
              \caption{ An example of the time-latitude diagram (same as in Figure~\ref{mapprof}a) being decomposed into empirical orthogonal functions (EOFs): (a) the primary data matrix as the window selected in the time-latitude diagram; (b) reconstruction of the primary data matrix as the sum of the first three (principal) decomposition terms (EOFs); (c) the same for the sum of the ten main EOFs of the decomposition. 
                      }
   \label{eof310}
   \end{figure}
%%%%%%%%%%%%%%%%%%%%%%%%%%%%%%%%%%%%%%%%%%%%%%%%%%%%%%%%%%%%%%%%%%%%%%%%%%%%%%%%

Figure~\ref{pamir} shows a three-dimensional view of the same section of the time-latitude diagram  as in Figure~\ref{mapprof}. The alternation of flows of two polarities can be seen in the form of ridges and depressions, the tops of which correspond to positive, and the dips correspond to negative fields. It should be noted that a similar structure of polarity alternation was found  in \citep{vern}. In this work, performed using the data of synoptic maps of the photospheric magnetic field (NSO Kitt Peak), a value of $+1$ or $-1$ was assigned to each pixel depending on the sign of the magnetic field. The similarity of the results of the two works shows that the alternation of polarities indicates the predominance of the magnetic field of a certain sign and does not depend on the magnitude of the field.

For the analysis of cyclic variations in polarity, four temporal intervals were selected with the most distinct alternation of flows of positive and negative polarities. Each time interval  included phases of decline, minimum and rise of solar activity cycle (from one RTTP to the other). Four intervals of time (two in the southern hemisphere and two in the northern), during which the polarity of the flows periodically changed, amounted to a total of about 40 years. It should be noted that during three of the four time intervals the variations were observed in the hemisphere with the positive polar field.
To study the cyclic structure of the magnetic field distribution, the method of empirical orthogonal functions (EOF) was used. A detailed description of the method and interpretation of the results obtained is given in the manual \citep{vene}.
Following this manual the analyzed data matrices $F$ were defined as windows cut from the time-latitude diagram. Four windows were selected with a pronounced cyclic change in the polarity of the magnetic field. 

Each of the windows was located between two RTTPs and occupied a section including phases of decline, minimum and rise of the solar cycle. The medium lifetime of the variations was on the order of 9 years. The window sizes were $P\times Q$ pixels ($P \sim 120$, $Q \sim 60$), where $P$ is the Carrington rotation number and $Q$ is the number of latitudinal  rows of the matrix  (latitude in the sine scale).

%%%%%%%%%%%%%%%%%%%%%%%%%%%%%%%%%%%%%%%%%%%%%%%%%%%%%%%%%%%%%%%%%%%%%%%%%%%%%%%Figure4
\begin{figure}[h] 
   \centerline{\includegraphics[width=0.95\textwidth,clip=]{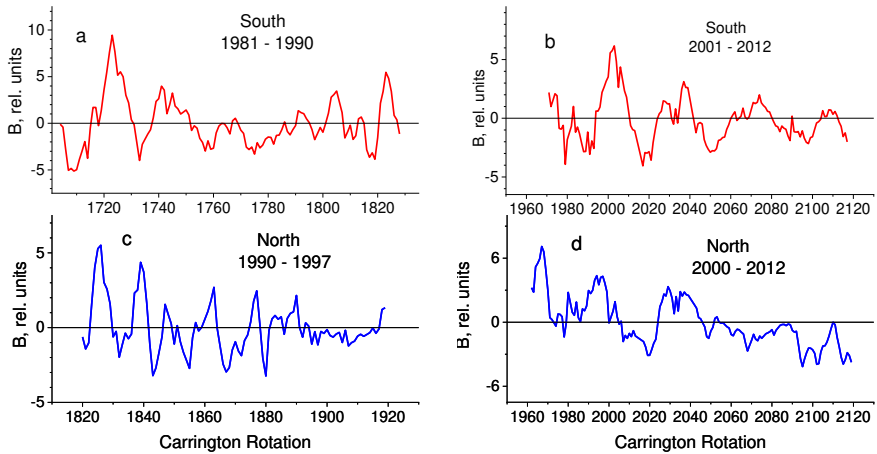}
              }
              \caption{ 
EOFs -- coefficients of decomposition into empirical orthogonal functions: first time functions for the 4 windows selected on the time-latitude diagram (a,b -- southern hemisphere; c,d -- northern hemisphere)
                      }
   \label{foureof}
   \end{figure}
%%%%%%%%%%%%%%%%%%%%%%%%%%%%%%%%%%%%%%%%%%%%%%%%%%%%%%%%%%%%%%%%%%%%%%%%

Time series (EOF expansion coefficients) were obtained for each of the windows in question. Briefly, the sequence of decomposition of the space-time matrix of data $F$ into eigenfunctions is as follows \citep{vene}. \\
1. Define the covariance matrix $R=F^t F$.\\
2. Find eigenvalues and eigenvectors of $R$.\\
3. Find the largest eigenvalues and their corresponding eigenvectors (EOFs).\\
4. Find the expansion coefficients by calculating $\vec{a}_j={F \cdot EOF_j}$

Typically, EOFs are represented in normalized form and ordered in ascending or descending eigenvalues. The corresponding decomposition factors should be arranged in the same way. Typically, several dominant eigenvalues determine the behavior of the data matrix. The ratio of the magnitude of an individual eigenvalue to the sum of all eigenvalues (in percent) indicates the contribution of the corresponding decomposition component to the total variation.
Primary data can be recovered using EOFs and decomposition coefficients according to the following formula:

\begin{equation}\label{eof}
    F=\sum_{j=1}^p \vec{a}_j \cdot EOF_j.
\end{equation}

Figure~\ref{eof310} shows an example of the reconstruction of the primary data matrix as sums of three and ten decomposition terms into EOF. As the initial data a section of the time-latitude diagram is used within latitudes from the equator to $-40^\circ$ of the southern hemisphere in the time interval 1981--1990. As the first eigenvalues dominate over the others, the most part of behavior of a matrix of data can be explained with only several eigenvectors which reflect the main features of primary data, diminishing influence of accidental noise. In this example, the largest relative eigenvalues for the window in question were 0.39, 0.20, and 0.13 ($72\%$ overall).
EOF decomposition confirms the presence of cyclic components in the distribution of magnetic fields. The first expansion coefficients are presented for the two S-hemisphere windows (Figure~\ref{foureof}a,b) and for the N-hemisphere (Figure~\ref{foureof}c,d). The contribution of the first decomposition component to the total variation for these four windows was on average $34\%$, the average period being 1.8 yrs.

\begin{figure}[t] 
   \centerline{\includegraphics[width=0.95\textwidth,clip=]{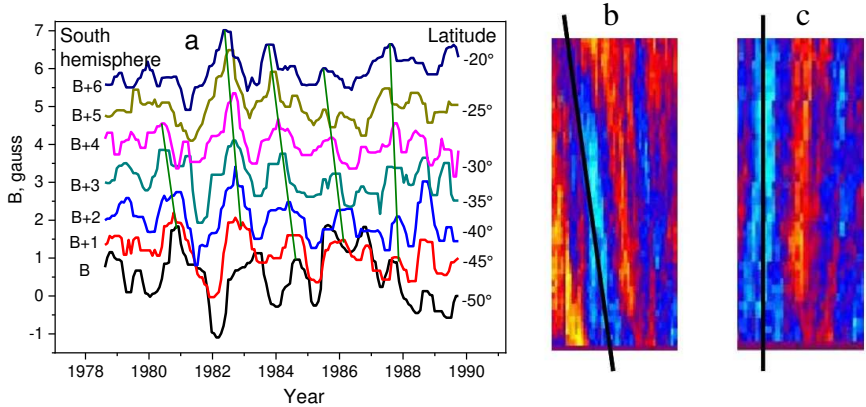}
              }
              \caption{ 
(a) Time profiles of the S-hemisphere magnetic field for a series of latitudes from $-20^\circ$  to $-50^\circ$. The profiles are shifted along the Y axis by 1\,G for easy viewing.
(b) Correction of the flow image prior to latitude averaging. On the left is a section of the time-latitude diagram with an inclined image of the flow, on the right is an adjusted section with the orientation of the image of the flow parallel to the Y axis.
                      }
   \label{shift}
   \end{figure}

The EOF method can be very useful as a tool for studying the cyclic patterns of the photospheric magnetic field. However, since the EOF method is not transparent enough, it is desirable to compare the EOF results with similar results obtained in simpler though less accurate, ways.
Due to fluctuations in the magnetic field, it is necessary to use time profiles averaged for the latitude interval for reliable isolation of cyclic components. However, with changes in latitude, there is a systematic shift in the pattern of alternating polarity accompanying the drift of magnetic flows towards the poles. Figure~\ref{shift}a shows the magnetic field time profiles for the latitudes of the southern hemisphere from $-20^\circ$ to $-50^\circ$ in $5^\circ$ increments. For convenience of viewing, the curves are shifted along the Y axis by 1\,G relative to each other. It can be seen that positions of the extrema of the curves change systematically following the increase of the latitude. The maxima and the minima of the curves correspond to positive and negative flows drifting towards the poles of the Sun.
%%%%%%%%%%%%%%%%%%%%%%%%%%%%%%%%%%%%%%%%%%%%%%%%%%%%%%%%Figure5
%%%%%%%%%%%%%%%%%%%%%%%%%%%%%%%%%%%%%%%%%%%%%%%%%%%%%%%%%%%%%%%%%%%%
%%%%%%%%%%%%%%%%%%%%%%%%%%%%%%%%%%%%%%%%%%%%%%%%%%%%%%%%%%%%%%%%%%%Figure6
\begin{figure}[h] 
   \centerline{\includegraphics[width=0.95\textwidth,clip=]{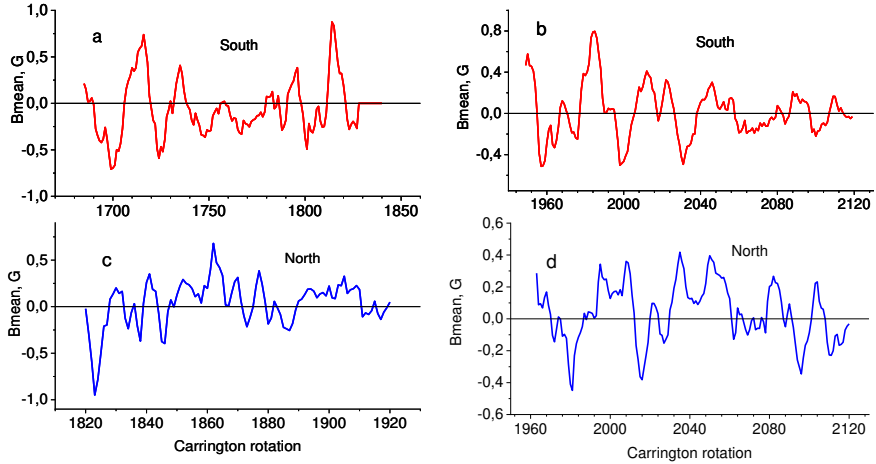}
              }
              \caption{ 
Extracting the cyclic component of the magnetic field for four areas of the time-latitude diagram.}
   \label{sums}
   \end{figure}
%%%%%%%%%%%%%%%%%%%%%%%%%%%%%%%%%%%%%%%%%%%%%%%%%%%%%%%%%%%%%%%%%%%%

The inclined straight lines connecting the maxima of the curves illustrate the latitudinal drift of positive flows with time.  In addition to this systematic movement the extrema of the curves in Figure~\ref{shift}a are subject to significant fluctuations. To reduce the effect of fluctuations in the analysis of magnetic flux parameters, time-latitude plots were averaged over latitude. When averaging the latitude of variations constructed for individual latitudes, a systematic shift must be taken into account (Figure~\ref{shift}a) by introducing an appropriate correction.
Before averaging, the diagram was adjusted so that the inclined image of the flow was oriented along the vertical axis. Figure~\ref{shift}b,c  shows two sections of the time-latitude diagram, with inclined magnetic flow image (b) and the same flow after correction of the diagram (c). The correction was carried out by selecting the optimal latitude offset to obtain the vertical direction of the flows. 

The time-latitude diagram is a matrix, each row of which may be interpreted as a time profile of a magnetic field at a fixed latitude.
After the above correction, the profiles were averaged for approximately the same four time intervals which were treated with the EOF method (Figure~\ref{foureof}). Figure~\ref{sums} shows the resulting time profiles for these four intervals: 1, 2 in the S-hemisphere, and 3, 4 in N-hemisphere. As the number of added profiles increases, fluctuations decrease, and the cyclic structure of the magnetic field appears more clearly. Figure~\ref{sums} shows the result of magnetic field averaging over the latitude intervals from $0^\circ$ to $+50^\circ$ and from $0^\circ$ to $-50^\circ$.
The following intervals were selected in the time-latitude diagram for the study of magnetic field cyclic variations: mean magnetic field  was calculated for 1980--1991 and 1999--2012 in the S-hemisphere  (Figure~\ref{sums}a,b); the N-hemisphere  variations are shown in Figure~\ref{sums}c,d  for 1990--1997 and 2000--2012. Each of these intervals covers the time between two adjacent RTTPs. The averaged dependences are used to estimate the period of magnetic field variations.

\begin{figure}[h] 
   \centerline{\includegraphics[width=0.75\textwidth,clip=]{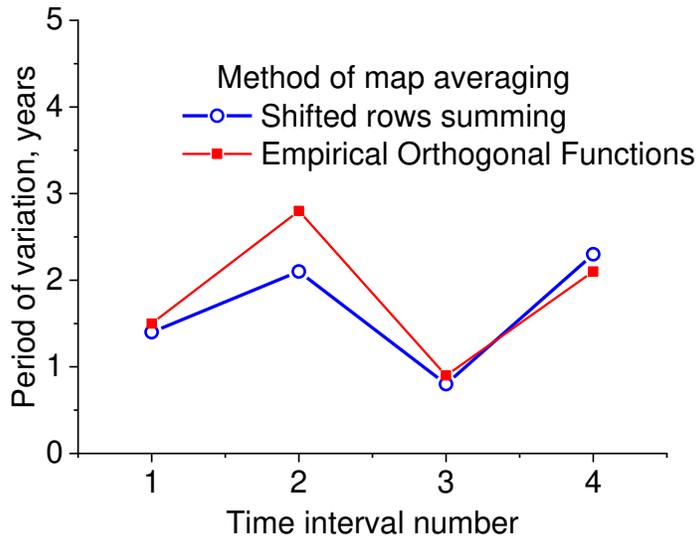}
              }
              \caption{ Evaluation of the periods of polarity alternation by two methods: decomposition of magnetic field into empirical orthogonal functions (EOFs) and averaging of the magnetic field time profiles     over the series of latitudes.
                      }
   \label{periods}
   \end{figure}

The results obtained by EOF and latitude summation (after image correction) are shown in Table~\ref{methods}, where the periods of variation obtain by two methods are presented for the four time intervals (1, 2 for the southern hemisphere and 3, 4 for the northern hemisphere).  The variation period varies from 0.8 to 2.8 years. The EOF average period value is 1.8 yr and the latitude summation method gives an average period of 1.65 yr.
The periods of variation determined by the two methods (latitude averaging and EOF decomposition) are compared in Figure~\ref{periods}. For four time intervals (two intervals in each of the hemispheres), similar results are obtained. The observed period values indicate a possible association of the sign alternation of flows with quasi biennial variations (QBO). The difference in variation 
periods  between  time intervals considered appears to be due to the features of each of the four intervals in question, rather than the random spread of results.

%%%%%%%%%%%%%%%%%%%%%%%%%%%%%%%%%%%%%%%%%%%%%%%%%%%%%%%%%%
\begin{table}[]
%\label{methods}
\centering
\begin{tabular}{clcc}
\hline
\multicolumn{2}{c}{Considered data}             & \multicolumn{2}{l}{Variation period } \\
\multicolumn{1}{l}{Hemisphere} & Time period    & T*, yrs                & T**, yrs                \\
\hline 
S                              & 1981.1 -- 1990.3 & 1.4                 & 1.5                 \\
S                              & 2001.1 -- 2012.0 & 2.1                 & 2.8                 \\
N                              & 1989.8 -- 1997.2 & 0.8                 & 0.9                 \\
N                              & 2000.4 -- 2012.1 & 2.3                 & 2.1   \\
\hline            
\end{tabular}
\caption{Variation periods obtained by two different methods: (a) by averaging of magnetic field time profiles for a series of latitudes (T*) and
(b) by decomposition into empirical orthogonal functions (T**).}
\label{methods}
\end{table}
%%%%%%%%%%%%%%%%%%%%%%%%%%%%%%%%%%%%%%%%%%%%%%%%%%%%%%%%%%

\section{Conclusions}\label{Concl}

Flows of magnetic fields with alternating polarity in the photosphere of the Sun has been studied. The data processing technique used made it possible to distinguish the features of the distribution of weak magnetic fields on the surface of the Sun. In weak fields, one of the two polarities alternately dominated in the form of flows about 1 year wide drifting from the equator to the poles. Flows were present on the Sun during the phase of decline, minimum and rise of the solar cycle, from one RTTP to another, for an average of 9 years. This period corresponds to the age of polar field constancy, with polarity variations appearing especially distinctly in a hemisphere with a positive field sign.
%%%%%%%%%%%%%%%%%%%%%%%%%%%Figure7
%%%%%%%%%%%%%%%%%%%%%%%%%%%%%%%%

To distinguish cyclic variations and study their parameters, a method of empirical orthogonal functions was used. It was shown that in each of the four windows of the time-latitude diagram with dimensions of $50^\circ$ in latitude and 120 Carrington rotations  in time there was a cyclic variation with an average period of 1.8 years. This variation accounted for 34\% of the total magnetic field variation. Evaluation of the period of variation using another method -- the averaging of the time-latitude data in the latitude interval of $50^\circ$ -- gave a close result of 1.65 years. These period values indicate a possible association of the observed variation with the other QBO manifestations. The obtained results indicate the presence of magnetic field flows in the photosphere with alternating dominant polarity. These flows have a number of stable characteristic features and appear, or disappear, in connection with the change in the sign of the polar field.

%%%%%%%%%%%%%%%%%%%%%%%%%%%%%%%%%%%%%%%%%%%%%%%%%%%%%%%%%%%%%%%%%%%%%%%%%%%
%% Acknowledgements
%
 \section*{Acknowledgements}

  The NSO/Kitt Peak data used here are produced cooperatively by NSF/NOAO, NASA/GSFC, and NOAA/SEL (ftp://nispdata.nso.edu/kpvt/synoptic/mag/). Data of the SOLIS instruments were provided by NISP/NSO/AURA/NSF, https://magmap.nso.edu/solis/archive.html

\medskip

%%% %%%%%%%%%%%%%%%%%%%%%%%%%%%%%%%%%%%%%%%%%%%%%%%%%%%%%%%%%%%\widetilde{\partial\partial}

%\end{article}

\end{document}